\documentclass[11pt]{article}
\usepackage{rotating,amsbsy,pifont,amssymb,bm,epsfig,amsmath}

\renewcommand\baselinestretch{1.10}
\def\br{{\mathbf r}} 
\def\bhr{\hat{\mathbf r}} 
\def\rG{{\mathrm G}} 
\def\rW{{\mathrm W}} 
\def\rP{{\mathrm P}} 
\def\rQ{{\mathrm Q}} 
\def\rM{{\mathrm M}}
\def\rV{{\mathrm V}}

\def\rZ{{\mathrm Z}} 
\def\rg{{\mathrm g}} 

\begin{document}

\title{Relativistic central--field Green's functions for the R{\large ATIP} 
       package}

\author{Peter Koval\footnote{Email: kovalp@physik.uni-kassel.de, 
        Tel: +49-561-804-4571, Fax: +49-561-804-4006.}
        \: and Stephan Fritzsche
        \\
        \\
        Fachbereich Naturwissenschaften, Universit\"a{}t Kassel,      \\
        Heinrich--Plett--Str 40,\  D--34132 Kassel, Germany.   
        }

\maketitle

\date{}

\setlength{\parindent}{0.5cm}

\begin{abstract}
From perturbation theory, Green's functions are known for providing a simple
and convenient access to the (complete) spectrum of atoms and ions. Having 
these functions available, they may help carry out perturbation 
expansions to any order beyond the first one. For most realistic potentials,
however, the Green's functions need to be calculated numerically since an
analytic form is known only for \textit{free} electrons or for their motion 
in a pure Coulomb field. Therefore, in order to facilitate the use of Green's 
functions also for atoms and ions other than the hydrogen--like ions, here 
we provide an extension to the \textsc{Ratip} program which supports the
computation of relativistic (one--electron) Green's functions in an --- 
arbitrarily given --- central--field potential $\rV(r)$.
Different computational 
modes have been implemented to define these effective potentials and
to generate the radial Green's functions for all bound--state 
energies $E\,<\,0$. In addition, care has been taken to provide a
user--friendly component of the \textsc{Ratip} package by utilizing
features of the Fortran 90/95 standard such as data structures,
allocatable arrays, or a module--oriented design.
\end{abstract}

\newpage
\setlength{\parindent}{0.0cm}

{\large\bf PROGRAM SUMMARY}

\bigskip

{\it Title of program:} \textsc{Xgreens}.

\bigskip

{\it Catalogue number:} To be assigned.

\bigskip

{\it Program obtainable from:} CPC Program Library, 
     Queen's University of Belfast, N. Ireland. 
     Users may also down--load a tar--file of the program 
     \texttt{ratip00.tar} from our home page at the University of Kassel
     (http://www.physik.uni-kassel.de/fritzsche/programs.html).

\bigskip
     
{\it Licensing provisions:} None.

\bigskip

{\it Computer for which the new version has been tested:}
     PC Pentium II, III, IV, Athlon.\newline
     {\it Installations:} University of Kassel (Germany).   \newline
     {\it Operating systems:} SuSE Linux 8.2, SuSE Linux 9.0. 
     
\bigskip
     
{\it Program language used in the new version:} ANSI standard Fortran 90/95.

\bigskip

{\it Memory required to execute with typical data:} On a standard grid 
     (400 nodes), one central--field Green's function requires about 50 kBytes 
     in RAM while approximately 3 MBytes are needed if saved as 
     two--dimensional array on some external disc space. 
     
\bigskip

{\it No.\ of bits in a word:} Real variables of double- and 
quad-precision are used.

\bigskip

{\it Peripheral used:} Disk for input/output.

\bigskip

{\it CPU time required to execute test data:}  2 minutes on a 450 MHz 
                                               Pentium III processor.
\bigskip

{\it Distribution format:} compressed tar file. 

\bigskip

Green's function, Kummer function, 
     multi-configuration Dirac--Fock, regular and irregular solutions.

\bigskip
     
{\it Nature of the physical problem:}  
     In atomic perturbation theory, Green's functions may help carry out the
     summation over the complete spectrum of atom and ions, including the 
     (summation over the) bound states as well as an integration over the 
     continuum [1]. Analytically, however, these functions are known only for 
     \textit{free} electrons ($ \rV (r) \:\equiv\: 0$) and for electrons in 
     a \textit{pure} Coulomb field ($\rV (r) \:=\: -Z/r$). For all other 
     choices of the potential, in contrast, the Green's functions must be
     determined numerically.

\bigskip

\textit{Method of solution:}  
     Relativistic Green's functions are generated for an arbitrary 
     central--field potential $\rV(r) \:=\: -\,\rZ (r)/r$ by using a piecewise 
     linear approximation of the effective nuclear charge function $\rZ(r)$ on
     some grid $r_i \;\, (i \,=\, 1, ..., N)$: 
     $\rZ_i(r)\,=\,Z_{0i}\,+\,Z_{1i}\,r$. Then, following McGuire's 
     algorithm [2], the radial Green's functions are
     constructed from the (two) linear--independent solutions of the 
     homogeneous equation [3]. In the computation of these radial functions, 
     the Kummer and Tricomi functions [4] are used extensively.
     
\bigskip

{\it Restrictions onto the complexity of the problem:}  
     The main restrictions of the program concern the shape of the effective
     nuclear charge $\rZ(r) \:=\: -r \,\rV(r)$, i.~e.\ the choice of the
     potential, and the allowed energies. Apart from obeying the proper 
     boundary conditions for a point--like nucleus, namely, 
     $\rZ(r\rightarrow 0)\, =\, Z_{\mathrm{nuc}}>0$ and 
     $\rZ(r\rightarrow \infty)\, =\, Z_{\mathrm{nuc}} \,-\, 
      N_{\mathrm{electrons}} \,\ge\, 0$, the first derivative of the charge
     function $\rZ(r)$ must be smaller 
     than the (absolute value of the) energy of the Green's function,
     $\frac{\partial \rZ(r)}{\partial r} \,<\, |E|$.
     
\bigskip
    
{\it Unusual features of the program:}  
     \textsc{Xgreens} has been designed as a part of the \textsc{Ratip} 
     package [5] for the calculation of relativistic atomic transition and 
     ionization properties. In a short dialog at the beginning of the
     execution, the user can specify the choice of the potential as well as
     the energies and the symmetries of the radial Green's functions to be
     calculated. Apart from central--field Green's functions, of course, 
     the Coulomb Green's function [6] can also be computed by selecting a 
     constant nuclear charge $\rZ (r)\,=\,Z_{\mathrm{eff}}$.
     In order to test the generated Green's functions, moreover, 
     we compare the two lowest bound--state orbitals which are calculated
     from the Green's functions with those as generated separately for the
     given potential. Like the other components of the \textsc{Ratip}
     package, \textsc{Xgreens} makes careful use of the Fortran 90/95 standard.

\bigskip

{\it References:}   \newline
     [1] R.~A.~Swainson and G.~W.~F.~Drake, J.~Phys.~A \textbf{24} (1991) 95.
     \newline  
     [2] E.~J.~McGuire, Phys.~Rev.~A \textbf{23} (1981) 186.
     \newline
     [3] P.~Morse and H.~Feshbach, Methods of Theoretical Physics
     (McGraw--Hill, New York 1953), \newline \hspace*{0.55cm}Part.~1, p.~825.
     \newline
     [4] J.~Spanier and B.~Keith, An Atlas of Functions 
     (Springer, New York, 1987).
     \newline
     [5] S.~Fritzsche, J.~Elec.\ Spec.\ Rel.\ Phen.\ \textbf{114--116} 
     (2001) 1155.
     \newline
     [6] P.~Koval and S.~Fritzsche, 
     % Relativistic wave and Green's functions for hydrogen--like ions,
     Comput.\ Phys.\ Commun. \textbf{152} (2003) 191.

\newpage
{\large\bf LONG WRITE--UP}

\bigskip

\setlength{\parindent}{0.5cm}

\section{\label{s:introduction}Introduction}

The use of Green's function has a long tradition for solving physical problems,
both in classical and quantum physics. From perturbation theory, for instance, 
these functions are known for providing a rather simple access to the 
complete spectrum of a quantum system and, hence, to facilitate the 
computation of perturbation expansions beyond the first order in perturbation 
theory. Applications of the Green's functions can be therefore found not only 
in atomic and molecular physics but also in quantum optics, field theory,
solid--state physics, and at various place elsewhere.

\medskip

In atomic physics, however, the use of Green's function methods was so far
mainly restricted to describe the motion of electrons in a pure Coulomb 
field, i.~e.\ to the theory of hydrogen and hydrogen--like ions.
For these one--electron systems, calculations have been carried 
out, for example, for the two--photon \cite{Laplancheetal:1976,Kovaletal:2003} 
and multi--photon ionization \cite{Gontier-and-Trachin:1973,Maquetetal:1998},
the two--photon decay \cite{Goldman-and-Drake:1981}, the second--order
contributions to the atomic polarizabilities \cite{Szmytkowski:1997}
as well as for determining radiative corrections 
\cite{Mohretal:1998, Indelicato-and-Mohr:2001}.

\medskip

Less attention, in contrast, has been paid to utilize Green's functions for
non--Coulomb fields or for describing the properties of many--electron atoms and
ions. Unlike to the generation of the --- bound and free--electron --- wave
functions to the Schr\"o{}dinger and Dirac equation, for which a number of
programs are available now also within the CPC--library
\cite{Parpiaetal:1996, Salvatetal:1995}, there is almost no code freely 
available which helps generate the relativistic central--field  Green's 
functions. As known from the literature, however, nonrelativistic 
central--field Green's functions were constructed by McGuire 
\cite{McGuire:1981} and by Huillier and coworkers \cite{Huillier:87}, and 
were successfully applied for studying the multi--photon ionization of 
valence--shell electrons in alkali atoms. Therefore, in order to facilitate 
the use of relativistic central--field Green's functions for atomic 
computations, here we describe and provide an extension to the 
\textsc{Ratip} package which calculates these functions (as the solution of 
the Dirac equation with a $\delta-$like inhomogeneity) for an arbitrary 
central field $\rV(r)$.

\medskip

In the following section, we start with summarizing the basic formulas
for the computation of relativistic central--field Green's functions. Apart 
from a brief discussion of the Dirac Hamiltonian, this includes the 
\textit{defining equation} for central--field Green's function and the 
separation of the three-dimensional Green's function into radial and
angular parts. However, since the separation has been discussed in detail
elsewhere in the literature \cite{Swainson-and-Drake:91:1-2-3}, we restrict
ourselves to a short account on that topic and mainly focus on the computation
of the radial components of the Green's functions. Section 
\ref{s:program-structure} later describes the program structure of 
\textsc{Xgreens}, its interactive control and how the code is distributed. 
Because the \textsc{Xgreens} program is designed as part of the \textsc{Ratip} 
package, we have used and modified several modules which were published before
along with other components of the program. Section \ref{s:test-calculations} 
explains and displays two examples of \textsc{Xgreens}, including (a) a dialog
in order to calculate a central--field potential from \textsc{Grasp92} wave 
functions \cite{Parpiaetal:1996} and (b) the generation of the 
radial Green's functions if the potential is loaded from an external file.
Finally, a short summary and outlook is given in section
\ref{s:summary-and-outlook}.

\medskip

\section{Theoretical background}

\subsection{Relativistic central--field Green's functions}

Most naturally, the relativistic central--field Green's function can be 
considered as a generalization of the (relativistic) Coulomb Green's 
function if, in the Dirac Hamiltonian\footnote{Here and in the following, 
we use atomic units ($m_e\,=\,\hbar\,=\,e^2/4\pi\epsilon_0\,=\,1$) if not 
stated otherwise.}
\begin{eqnarray} 
\label{DH}
   \hat H_{\rm D}\,(\br) & = &  
   -i c \bm{\alpha} \,\nabla \,+\, \beta c^2 \,+\, \rV(r)  \, ,
\end{eqnarray}
the Coulomb potential is replaced by some (arbitrarily given)
central--field potential,
$\rV^{\rm C} (r)\,=\,-\frac{Z}{r} \:\longrightarrow\: -\frac{\rZ(r)}{r}$.
As in the nonrelativistic case where the Green's function obeys the
\textit{defining} equation
$\: (\,\hat H\,-\,E\,)\,\rG_E(\br,\,\br')\,=\,\delta(\br\,-\,\br') \:$, the 
relativistic Green's function is given by a $4\,\times\,4$ matrix 
\cite{DrakeHandBook} which satisfies the inhomogeneous equation
\begin{eqnarray} 
\label{CoulombDiracGreensEquation} 
   \left( \hat H_{\rm D} (\br)-E - c^{\,2}\, \right) \, \rG_E(\br,\br') 
   & = & \delta(\br-\br') \: \mathbf{I}_{\,4} \, ,
\end{eqnarray}
with $\mathbf{I}_{\,4}$ being the $4\times 4$ unit--matrix and where, 
as usual in atomic structure theory, $E$ refers to the total energy of the
electron but without its rest energy $m_e\,c^{\,2}$. Moreover, since in polar 
coordinates a central--field potential $\rZ(r)$ in the Hamiltonian (\ref{DH}) 
does not affect the separation of the variables, the central--field Green's
function has the same \textit{radial--angular} representation as in the pure
Coulomb case \cite{Swainson-and-Drake:91:1-2-3}

\begin{small}
\begin{eqnarray} 
\label{greens_radial_Green_matrix} 
   \rG_E(\br,\br') & = & \sum\limits_{\kappa m}\frac{1}{rr'} 
   \begin{pmatrix} 
      \rg_{E\kappa}^{\,LL}\,(r,\,r')\: 
      \Omega_{\kappa m}\,(\bhr)\, \Omega_{\kappa m}^{\dagger}(\bhr') & 
      -i\,\rg_{E\kappa}^{\,LS}\,(r,\,r')\: 
      \Omega_{\kappa m}\,(\bhr)\, \Omega_{-\kappa m}^{\dagger}(\bhr')\: 
      \\[0.25cm] 
      \,i\,\rg_{E\kappa}^{\,SL}\,(r,\,r')\: 
      \Omega_{-\kappa m}\,(\bhr)\,\Omega_{\kappa m}^{\dagger}(\bhr') & 
      \rg_{E\kappa}^{\,SS}\,(r,\,r')\: 
      \Omega_{-\kappa m}\,(\bhr)\,\Omega_{-\kappa m}^{\dagger}(\bhr') 
   \end{pmatrix} \, ,
\label{cfgf-separation}
\end{eqnarray}
\end{small}
\\[-0.1cm]
where $\Omega_{\kappa m}(\bhr) \,=\, \Omega_{\kappa m} (\vartheta,\varphi)$ 
denotes a standard spherical Dirac--spinor and 
$\kappa \,=\, \pm\, (j+1/2)\,$ for $\,l \,=\, j \pm 1/2\,$ is the 
\textit{relativistic} angular momentum quantum number; this number carries 
information about both the total angular momentum $j$ as well as the 
parity $(-1)^l$ of the Green's function. While the summation over 
$\: \kappa \,=\, \pm 1, \, \pm 2, \, ... \;$ runs over all (non--zero) 
integers, the summation over the magnetic quantum number 
$\:m \,=\, -j,\, -j+1,\, \ldots,\, j\:$ is restricted by the corresponding
total angular momentum. Moreover, the radial part of the central--field 
Green's function
{\small$\begin{pmatrix}
    \,\rg_{E\kappa}^{\,LL}\,(r, r') & \rg_{E\kappa}^{\,LS}\,(r, r') \\[0.2cm] 
    \,\rg_{E\kappa}^{\,SL}\,(r,r') & \rg_{E\kappa}^{\,SS}\,(r,r')\, 
 \end{pmatrix}$} in (\ref{greens_radial_Green_matrix}) can be treated simply
as a $2\times 2$ matrix function which satisfies the equation

\begin{small}
\begin{eqnarray} 
   \begin{pmatrix} 
   \left[\displaystyle - \frac{\alpha \,\rZ(r)}{r} - \alpha E \right]  &  
   \left[\displaystyle \frac{\kappa}{ r} - \frac{\partial}{\partial r} \right]
    \\[0.35cm] 
   \left[\displaystyle \frac{\partial}{\partial r}  
         + \frac{\kappa }{ r} \right]                     & 
   \left[\displaystyle -\frac{2}{\alpha}
         -\frac{\alpha\, \rZ(r)}{r} - \alpha E \right] 
   \end{pmatrix} \, 
   \begin{pmatrix}\rg_{E\kappa}^{LL} & \rg_{E\kappa}^{LS} \\[0.2cm] 
      \rg_{E\kappa}^{SL} & \rg_{E\kappa}^{SS} 
   \end{pmatrix} & = & \alpha\, \delta(r-r')\, \mathbf{I}_2 \, ,
\label{rgf-first-equation}
\end{eqnarray}
\end{small}
\\[-0.1cm]
with $\mathbf{I}_{\,2}$ now being the $2\times 2$ unit--matrix. Note that 
in Eq.\  (\ref{rgf-first-equation}), $\alpha$ refers to Sommerfeld's 
fine structure constant and that, in order to keep the equations similar to 
those for the Coulomb Green's functions \cite{Swainson-and-Drake:91:1-2-3},
we make use of a nuclear charge function $\rZ(r)\,=\,-r \rV(r)$
to define the central--field potential, instead of $\rV (r)$ explicitly. 
In the radial--angular representation (\ref{greens_radial_Green_matrix}),
two superscripts $T$ and $T'$ were introduced to denote the individual components
in the $2\times 2$ radial Green's function matrix. These superscripts take 
the values $\,T\,=\,L\,$ or $\,T\,=\,S\,$ to refer to either the \textit{large} 
or \textit{small} component, respectively, when multiplied with a corresponding 
(2--spinor) radial solution of the Dirac Hamiltonian (\ref{DH}). 
For a pure Coulomb potential, $ \rZ(r) \,\equiv\, Z_{\mathrm{eff}}$, an 
explicit representation of the (four) components  
$ \rg^{\,TT'}_{E\kappa}\,(r, \, r')$ of the radial Green's function can be 
found in Refs.\  \cite{Swainson-and-Drake:91:1-2-3,DrakeHandBook,
Koval-and-Fritzsche:2003}.

\subsection{\label{ss:generation-of-rgf}Generation of radial Green's functions}

When compared with the Coulomb Green's functions, not much need to be changed 
for the central--field functions in Eqs.~(\ref{cfgf-separation}) and 
(\ref{rgf-first-equation}) except that the nuclear charge $ \rZ \,=\, \rZ (r)$
now depends on $r$ and, hence, that \textit{analytic solutions} to these
equations are no longer available. Therefore, to find a numerical solution 
to Eq.\  (\ref{rgf-first-equation}), let us first mention that this matrix 
equation just describes \textit{coupled} equations for the two independent 
pairs $(\rg^{\,LL}_{E\kappa},\, \rg^{\,SL}_{E\kappa})$ and 
$(\rg^{SS}_{E\kappa}, \,\rg^{LS}_{E\kappa})$ of the radial components. 
For example, if we consider the first pair 
$(\rg^{\,LL}_{E\kappa},\, \rg^{\,SL}_{E\kappa})$ of radial components of 
the Green's function, it has to satisfy the two equations
\begin{eqnarray} 
\label{one-electron-radial-equation-first-couple-first}   
   \left[\displaystyle - \frac{\alpha\, \rZ(r)}{r} - \alpha E \right]
   \, \rg_{E\kappa}^{\,LL} \, (r,\,r') +
   \left[\displaystyle \frac{\kappa}{ r} -
   \frac{\partial}{\partial r} \right]
   \, \rg_{E\kappa}^{\,SL} \, (r,\,r') & = & \alpha \,\delta(r-r'), 
   \\[0.2cm]
\label{one-electron-radial-equation-first-couple-second}
   \left[\displaystyle \frac{\partial}{\partial r}  
         + \frac{\kappa }{ r} \right] \, \rg_{E\kappa}^{\,LL} \, (r,\,r')\, +\, 
    \left[\displaystyle -\frac{2}{\alpha}- 
    \frac{\alpha\, \rZ(r)}{r} - \alpha E \right]\,
    \rg_{E\kappa}^{\,SL} \, (r,\,r') & = & 0 \, ,
\end{eqnarray}
and a similar set of equations holds for the second pair
$(\rg^{\,SS}_{E\kappa}, \,\rg^{\,LS}_{E\kappa})$. 
In the following, therefore, we will discuss the algorithm for solving Eqs.\  
(\ref{one-electron-radial-equation-first-couple-first}) and 
(\ref{one-electron-radial-equation-first-couple-second}) but need not display
the analogue formulas for the pair 
$(\rg^{\,SS}_{E\kappa}, \,\rg^{\,LS}_{E\kappa})$.

\medskip

Inserting $\rg^{\,SL}_{E\kappa}\, (r,\,r')$ from 
Eq.\  (\ref{one-electron-radial-equation-first-couple-second}) into
(\ref{one-electron-radial-equation-first-couple-first}), we arrive at the 
second--order inhomogeneous differential equation
\begin{equation}
\label{one-electron-radial-equation-gll}
   \left(\frac{\partial }{\partial r}-\frac{\kappa}{r}\right) %\left(
   \frac{\left(\displaystyle
   \frac{\partial}{\partial r}+
         \frac{\kappa}{r}\right)\,\rg_{E\kappa}^{\,LL}(r,r')}
   {\displaystyle \frac{2}{\alpha} + 
    \frac{\alpha \,\rZ(r)}{r}+ \alpha E } %\right)
   \,+ \,\left(\frac{\alpha \,\rZ(r)}{r} + 
         \alpha \,E\right) \, \rg_{E\kappa}^{\,LL}(r,r')
   \:=\:\alpha \,\delta (r-r') \, 
\end{equation}
for the component $\rg^{\,LL}_{E\kappa}(r,\,r')$. Solution of this equation 
can be constructed as product of two linearly independent solutions 
\cite{DrakeHandBook}
\begin{equation}
\label{product-M-W}
   \rg_{E\kappa}^{\,LL} \,(r, r') \:=\: 
   \rM_{E\kappa}^{\,LL} \,(\min(r,r'))\:\cdot\:
   \rW_{E\kappa}^{\,LL} \,(\max(r,r'))
\end{equation}
for the corresponding homogeneous case [cf.\ 
Eq.~(\ref{one-electron-radial-homogeneous-equation}) below], where
$\rM_{E\kappa}^{\,LL} \,(r)$ denotes a solution which is regular at the 
origin, and $\rW_{E\kappa}^{\,LL} \,(r)$ a solution regular at infinity. Below, 
we will obtain these functions following the numerical procedure as 
suggested by McGuire \cite{McGuire:1981}.

\medskip

For this, let us start with approximating the nuclear charge function 
$\rZ (r)$ on some grid $\,r_i, \;\, i\,=\,1,\,\ldots,\, i_{\mathrm{max}}\:$
in terms of a set of straight lines
\begin{equation}
\label{piecewise-linear-approximation}
   \rZ_{\,i} (r) \,=\, Z_{\,0i} \,+\, Z_{\,1i}\, r, \qquad
   \text{ for }  \quad r_{i} \,\le\, r \,\le\, r_{i+1} \, ,
\end{equation}
and from the \textit{homogeneous part} of Eq.\  
(\ref{one-electron-radial-equation-gll})
\begin{equation}
\label{one-electron-radial-homogeneous-equation}
   \left(\frac{\partial }{\partial r}-\frac{\kappa}{r}\right)%\left(
   \frac{\left(\displaystyle
   \frac{\partial}{\partial r}+\frac{\kappa}{r}\right)\,
   \rg_{E\kappa}^{\,i}\,(r)}
   {\displaystyle \frac{2}{\alpha}  + 
    \frac{\alpha \,\rZ_{\,i}(r)}{r} + \alpha \,E } %\right)
   \,+\, \left(\frac{\alpha\, \rZ_{\,i} (r)}{r}+\alpha \,E\right) \,
   \rg_{E\kappa}^{\,i} \,(r) \:=\:0 
\end{equation}
as given for the $i$--th interval $ [r_i,\, r_{i+1}]$ of the grid. 
In this Eq., we can drop the second argument $r'$ in the component 
$\rg_{E\kappa}^{\,LL}$, since it now appears only as a parameter, and
replace it by the superscript $i$ to denote the particular piece of the grid
for which we want the solution. From the approximation
(\ref{piecewise-linear-approximation}), moreover, we see that the 
nuclear charge function $\rZ_{\,i} (r)$ within the $i-$th interval
gives rise to a \textit{pure} Coulomb potential $\rZ_{\,0i} / r$
and a constant which is simply \textit{added} to the energy: 
$\,E \,\rightarrow\, \rZ_{\,0i} \,+\, E$. We therefore find 
that, within the given interval, the regular and 
irregular solutions at the origin in (\ref{product-M-W}) can both be written
as linear combinations of the corresponding solutions for a Coulomb field
\begin{eqnarray}
\label{piecewise-M}
   \rM_{E\kappa}^{\,i} (r) &=& 
   f_{i,1}\,\rM_{E\kappa}^{\,i,{\rm\, Coulomb}} (r) \,+\,
   f_{i,2}\,\rW_{E\kappa}^{\,i,{\rm\, Coulomb}} (r)       \\[0.2cm]
\label{piecewise-W}
   \rW_{E\kappa}^{\,i} (r) &=& 
   g_{i,1}\,\rM_{E\kappa}^{\,i,{\rm\, Coulomb}} (r) \,+\,
   g_{i,2}\,\rW_{E\kappa}^{\,i,{\rm\, Coulomb}} (r) \, 
\end{eqnarray}
with constants $\{\,f_{i,1}, \,f_{i,2}, \,g_{i,1}, \,g_{i,2} \}$ which still
need to be determined. For the Coulomb potential, the functions
$\rM_{E\kappa}^{\,i,{\rm\, Coulomb}} (r)$ and 
$\rW_{E\kappa}^{\,i,{\rm\, Coulomb}} (r)$ are known analytically and given
by \cite{DrakeHandBook}
\begin{eqnarray}
\label{general-solution-m}
   \rM_{E\kappa}^{\,i,{\rm\, Coulomb}} (r) &=&
   r^{s_i} \, e^{-q_i r} \, \left[ t_i\,\rM(-t_i+1, 2s_i+1, 2q_ir) \right.
   \nonumber \\[0.1cm] 
   &  & \hspace*{3.0cm} \left.
   \, + \,(\kappa - Z_{0i}/q_i) \,\rM(-t_i, 2s_i+1, 2q_ir) \right], \\[0.25cm]
\label{general-solution-w}
   \rW_{E\kappa}^{\,i,{\rm\, Coulomb}} (r) &=&
   r^{s_i} \, e^{-q_i r} \, \left[(\kappa + Z_{0i}/q_i)\, 
   \mathrm{U}(-t_i+1, 2s_i+1, 2q_ir) \right.
   \nonumber \\[0.1cm] 
   &  & \hspace*{4.8cm} \left.
   \,+\,
   \mathrm{U}(-t_i, 2s_i+1, 2q_ir) \right] \, ,
\end{eqnarray}
where $\rM(a,\, b,\, r)$ and $\mathrm{U}(a,\, b,\, r)$ denote the 
Kummer and Tricomi functions \cite{Spanier&Keith, Abramowitz}, respectively,
and where the quantities $s_i$, $t_i$ and $q_i$ are given by
\begin{eqnarray}
\label{st-definition}
   s_i &=& \sqrt{\kappa^2 - \alpha^2 Z_{0i}^2} \, , \qquad\qquad
   t_i \;=\; \frac{\alpha Z_{0i} ((E+Z_{1i})\alpha^2+1)}{
                   \sqrt{1-((E+Z_{1i})\alpha^2+1)^2}} -s_i, \\[0.2cm]
\label{q-definition}
   q_i &=& \sqrt{-(E+Z_{1i})((E+Z_{1i})\alpha^2+2)} \; .
\end{eqnarray}

The set of constants $\{\,f_{i,1}, \,f_{i,2}, \,g_{i,1}, \,g_{i,2};\;
i \,= \, 1,\,...,\, i_{\rm max} \}$ can be determined from the fact 
that the two functions $\,\rM_{E\kappa}^{\,LL} (r)\,$ and 
$\,\rW_{E\kappa}^{\,LL} (r)\,$ 
in ansatz (\ref{product-M-W}) as well as their derivatives need to be 
continuous in $r$, and that they behave \textit{regularly} at the origin or 
at infinity, respectively. The constraint of being regular at the origin,
for instance, requires the coefficients $f_{i=1,1}\,=\,1$ and 
$f_{i=1,2}\,=\,0$ and can be used together with the continuity 
$\,\rM_{E\kappa}^{\,LL} (r)\,$ and $\,\rM_{E\kappa}^{\,\prime\,LL} (r)\,$,
\begin{eqnarray}
\label{continuity-equations-f}
   f_{i,1}\,   M_{E\kappa}^{\,i,{\rm\, C}}   (r_i)  \,+\,
   f_{i,2}\,   W_{E\kappa}^{\,i,{\rm\, C}}   (r_i) & = &
   f_{i+1,1}\, M_{E\kappa}^{\,i+1,{\rm\, C}} (r_i)  \,+\,
   f_{i+1,2}\, W_{E\kappa}^{\,i+1,{\rm\, C}} (r_i), \\[0.2cm]
\label{continuity-equations-f-prime}
   f_{i,1}\,   M_{E\kappa}^{\,\prime\, i,{\rm\,   C}} (r_i)  \,+\,
   f_{i,2}\,   W_{E\kappa}^{\,\prime\, i,{\rm\,   C}} (r_i) & = &
   f_{i+1,1}\, M_{E\kappa}^{\,\prime\, i+1,{\rm\, C}} (r_i) \,+\,
   f_{i+1,2}\, W_{E\kappa}^{\,\prime\, i+1,{\rm\, C}} (r_i).
\end{eqnarray} 
[where the superscript C here refers to the Coulomb functions in Eqs.\
(\ref{general-solution-m}) and (\ref{general-solution-w})] in order to 
determine all the coefficients $f_{i,j}$ up to a normalization constant. 
A similar recurrence procedure also applies to the coefficients 
$g_{i,j}$, but by starting from 'infinity', that is with 
$g_{i_{\rm max},1}\,=\,0$ and $g_{i_{\rm max},2}\,=\,1$, and by going 
\textit{backwards} in the index $i$ towards the origin.

\medskip

To determine finally the normalization of the radial component
$\,\rg^{\,LL}_{E\kappa}(r,\,r')\,$, e.~g.\ of the coefficients 
$\{ f_{i,j},\, g_{i,j} \}$, we may return to Eq.\  
(\ref{one-electron-radial-equation-gll}) and re--write it in the form
\begin{eqnarray}
\label{h0-form}
   \left( \hat{h}_0 \:-\: E \right) \, \rg^{\,LL}_{E\kappa}(r,\,r')
   & = & \delta (r-r') \, .
\end{eqnarray}
Taking the integral over latter equation, we see that 
$$
\int_{r'-\varepsilon}^{r'+\varepsilon}\,(\,\hat h_0\,-\,E\,)
\ \rg_{E\kappa}^{LL}(r, r')\; dr\, =\,
\int_{r'-\varepsilon}^{r'+\varepsilon}\, \delta(r-r') \, dr \, = \, 1 
$$
should be valid for $\,\varepsilon \,\rightarrow\, + \,0\,$ and for any
$r'$ and, hence, that --- by using Eq.\ (\ref{one-electron-radial-equation-gll})
and carrying out some algebraic manipulations --- the derivative 
\begin{equation}
\label{jump}
   \left. \frac{\partial }{\partial r} \: \rg_{E\kappa}^{\,LL} (r,\, r') \;
   \right|_{\,r\,=\,r'-\varepsilon}^{\,r\,=\,r'+\varepsilon} \;=\;
   \alpha \, \left( \frac{2}{\alpha} \,+\,
                    \frac{\alpha \, \rZ(r')}{r'} + \alpha E \right) \, 
\end{equation}
'jumps' at $r \,=\, r'$. We can use the right hand side of Eq.\ (\ref{jump})
together with the derivative of Eq.\  (\ref{product-M-W}) to determine the
normalization constants $c_f$ and $c_g$ with which the coefficients
$\, f_{i,j}\,$ and $\, g_{i,j}\,$ need to be multiplied in order to obtain
the normalized radial component $\,\rg^{\,LL}_{E\kappa}(r,\,r')\,$.

\medskip

Having constructed $\rg^{\,LL}_{E\kappa} \, (r,\,r')$ piecewise as solution 
to Eq.~(\ref{one-electron-radial-equation-gll}), we may obtain the second 
component $\rg^{\,SL}_{E\kappa} \,(r,\,r')$ of this pair simply from Eq.\
(\ref{one-electron-radial-equation-first-couple-second}),
\begin{equation}
\label{jump-SL}
   \rg^{\,SL}_{E\kappa} \,(r,\, r') \:=\: 
   \frac{\left(\displaystyle\frac{\partial}{\partial r} + 
   \frac{\kappa}{r}\right ) \rg^{\,LL}_{E\kappa} \,(r,\,r')}
   {\displaystyle \frac{2}{\alpha} + \frac{\alpha \,\rZ(r)}{r} + 
    \alpha \,E} \, ,
\end{equation} 
where the derivatives of the Kummer and Tricomi functions in expressions
(\ref{general-solution-m}) and (\ref{general-solution-w}) can be calculated 
by means of standard formulae \cite{Abramowitz}
\begin{eqnarray*}
   \mathrm{M}'(a,b,z) = \frac{a}{b}\,\mathrm{M}(a+1,b+1,z)\, , \ \ \quad
   \mathrm{U}'(a,b,z) = -a\,\mathrm{U}(a+1,b+1,z) \, .
\end{eqnarray*}
In practise, both functions $\mathrm{M}(a,b,z)$ and $\mathrm{U}(a,b,z)$ are
required only for real arguments $a$, $b$, and $z$ but need to be calculated
with a rather sophisticated algorithm \cite{Koval-PhD-thesis} in order to
ensure numerical stability and to provide sufficiently accurate results.
A similar numerical procedure has to be carry out also for the second pair
$(\rg^{\,SS}_{E\kappa}, \,\rg^{\,LS}_{E\kappa})$ of radial components.
This gives rise of course to another set of coefficients 
$\{ \, \tilde{f}_{i,j},\, \tilde{g}_{i,j} \}$ and, finally, to the
full radial Green's function from Eq.\  (\ref{rgf-first-equation}).
The $4\,i_{\rm\, max}$ 
coefficients for the (piecewise) regular and irregular 
solutions in ansatz (\ref{piecewise-M}--\ref{piecewise-W}) and 
(\ref{product-M-W}) certainly provide --- 
together with a few numerical procedures --- the most \textit{compact} 
representation of the radial central--field Green's functions. 
For the further computation of matrix elements and atomic properties, 
however, these radial components are usually represented (and stored) at 
some (2--dimensional) grid in $r$ and $r'$ as we will discuss below.

\subsection{\label{ss:accuracy-test}Tests on the accuracy of the radial 
            Green's functions}

To make further use of the Green's functions in applications, it is necessary
to have a simple test on their \textit{numerical} accuracy. For the 
radial--spherical representation of these functions as defined in 
Eq.~(\ref{cfgf-separation}), 
such a test is easily constructed since the Green's function contains the
information about the complete spectrum of the Hamiltonian (\ref{DH}) and,
hence, about any of its eigenstates $\psi_n (\br)$. Making use of the 
well--known expansion of the Green's function
$$  \rG_E(\br,\,\br')\,=\,
    \sum_{n}\frac{\psi_n(\br)\,\psi_n^{\dagger}(\br')}{E_{n}\,-\,E}
$$
in terms of the eigenstates of the Hamiltonian, the relation
\begin{equation}
\label{test-relation}
   \psi_n (\br') \, = \, 
   (E_n - E)\,\int\, d\br \: \psi_{n}^{\dagger}(\br)\,\rG_E(\br,\,\br')\, 
\end{equation}
can be easily derived from the orthogonality of the eigenfunctions and holds 
for all energies $E \,\ne\, E_{n}$. With the representation
(\ref{cfgf-separation}) in mind, the relation (\ref{test-relation}) can be 
written also in terms of the radial wave and Green's functions
\begin{equation}
\label{rgf-check}
   \begin{pmatrix} \tilde\rP_{n\kappa}\,(r') \\[0.2cm] 
                   \tilde\rQ_{n\kappa}\,(r') \end{pmatrix}
   \, = \, (E_{n\kappa} - E)\;\int_0^{\infty} \, dr \:
   \left( \rP_{n\kappa}(r), \, \rQ_{n\kappa}(r) \right)\,
   \begin{pmatrix}\rg_{E\kappa}^{\,LL} \,(r,\,r') & 
                  \rg_{E\kappa}^{\,LS} \,(r,\,r') \\[0.2cm] 
                  \rg_{E\kappa}^{\,SL} \,(r,\,r') & 
		  \rg_{E\kappa}^{\,SS} \,(r,\,r') 
   \end{pmatrix}  \, ,
\end{equation}
where $\rP_{n\kappa}\,(r)$ and $\rQ_{n\kappa}\,(r)$ are the large and small
components of the (Dirac) radial 2--spinors. As indicated by the 
tilde on the left--hand--side of Eq.\  (\ref{rgf-check}), therefore,
this relation may serve as a test on the accuracy of the Green's functions 
if, for instance, the radial components from both side of the equation 
are compared with each other or if some proper \textit{overlap} integral 
is calculated.

\medskip

To test on the precision of the Green's functions in \textsc{Xgreens}, 
we generate the radial bound--state wave function  $\rP_{n\kappa} (r)$ and 
$ \rQ_{n\kappa}(r)$ as solution of the radial Dirac equation 
\begin{eqnarray} 
   \left[\displaystyle - \frac{\alpha\, \rZ(r)}{r} - \alpha E \right]
   \, \rP_{n\kappa}(r) +
   \left[\displaystyle \frac{\kappa}{ r} -
   \frac{\partial}{\partial r} \right]
   \, \rQ_{n\kappa}(r) & = &0 
    \label{one-electron-radial-equation-first}   
   \\[0.1cm]
   \left[\displaystyle \frac{\partial}{\partial r}  
         + \frac{\kappa }{ r} \right] \, \rP_{n\kappa}(r)\, +\, 
    \left[\displaystyle -\frac{2}{\alpha}- 
    \frac{\alpha\, \rZ(r)}{r} - \alpha E \right]\,
    \rQ_{n\kappa}(r) & = &0
    \label{one-electron-radial-equation-second}
\end{eqnarray}
with the same potential $ - \frac{Z(r)}{r}$ as applied before and by making use
of the program by Salvat {{\it etal }} \cite{Salvatetal:1995}. This solver has been
embedded in our code and is utilized in order to calculate the relativistic 
components. With these radial functions, we then compute the two integrals in 
(\ref{rgf-check}) for obtaining $\tilde\rP_{n\kappa} \,(r)$ and 
$\tilde\rQ_{n\kappa} \,(r)$, respectively, as a function of $r$. 
As default in \textsc{Xgreens}, we use the \textit{overlap} integral
\begin{equation}
\label{overlap-integral}
   \int \, \left(\, \tilde\rP_{n\kappa}(r)\:\rP_{n\kappa}(r)+
                    \tilde\rQ_{n\kappa}(r)\:\rQ_{n\kappa}(r) \,
            \right) \; dr\, 
\end{equation}
for the two lowest principal quantum numbers $n$ of the (given) symmetry 
$\kappa$, together with the corresponding \textit{normalization} integral
\begin{equation}
   \int\,(\,\tilde\rP_{n\kappa}(r)\tilde\rP_{n\kappa}(r)+
   \tilde\rQ_{n\kappa}(r)\tilde\rQ_{n\kappa}(r)\,)\,\,dr\, ,
\end{equation}
as a \textit{numerical measure} on the accuracy of the generated Green's
functions. These integrals are displayed explicitly by the program (on demand).
For a standard (logarithmic) grid with about 300 nodes, the overlap
integral (\ref{overlap-integral}) is typically within the range
$10^{-2} \,\ldots\, 10^{-3}$. For such a grid, the accuracy is limited in the 
computations by the linear interpolation of the wave and Green's functions.
The accuracy can be increased however for a larger number $i_{\rm\, max}$ of
grid points as shown in section \ref{s:test-calculations}.

\section{\label{s:program-structure}Program structure}

\subsection{The R{\small ATIP} package}

Similar as \textsc{Grasp92} \cite{Parpiaetal:1996} was designed for generating
the wave functions within the multi-configuration Dirac--Fock (MCDF) model,
the \textsc{Ratip} package \cite{Fritzsche:01,Fritzsche/CFF/Dong:00,
Fritzscheetal-RELCI:02} is organized as a suite of program components
to calculate a variety of (relativistic) atomic transition and
ionization properties. Among other features, the components of \textsc{Ratip}
support for instance investigations on the autoionization of atoms and ions,
the interaction with the radiation field, the parametrization of angular 
distributions in the emission of electrons and photons, or the analysis of 
interference effects between radiative and non--radiative processes. 
In quite different case studies, \textsc{Ratip} has helped analyze and 
interprete a large number of spectra and experiments. In order to provide 
efficient tools for atomic computations, it also incorporates a number of 
further components (beside of the main program components for calculating 
certain physical properties) which facilitate the transformation of wave 
functions between different coupling schemes or the generation of continuum 
orbitals and angular coefficients. With the development of the 
\textsc{Xgreens} program, we now provide an additional component to generate 
the relativistic central--field Green's functions within the \textsc{Ratip} 
environment. This is rather independent of their later 'use' where different  
properties such as the two--photon ionization and decay or various
polarizabilities of atoms and ions might be calculated with the help of these
functions. Owing to the large number of possible applications, however, 
a support of certain properties will  largely depends on the requests by the 
users and on our further experience with the code.

\medskip

Not much more need to be said about \textsc{Ratip}'s overall structure.
A detailed account on its present capabilities has been given previously
\cite{Fritzsche:02}. With regard to the further development of \textsc{Ratip}, 
we just note that our main concern now pertains to a long life--cycle of 
the code and to an object--oriented design within the framework of 
Fortran 90/95. With the present set--up of the \textsc{Xgreens} component, 
we continue our effort for providing an atomic code which is prepared for 
future applications in dealing with open--shell atoms and ions.

\subsection{\label{s:internal-structures}Data structures and program execution}

Certainly, the major purpose of \textsc{Xgreens} is the computation of
(one--electron) radial Green's functions for some --- specified or externally
given --- central--field potential, which can later be utilized for 
calculating radial matrix elements of the type
\begin{equation}
   U_{\,\beta\,\kappa\,\alpha}^{\,T_{\beta} T \tilde{T} T_{\alpha}} \,
   (k,\Lambda;\,E;\,\tilde{k},\tilde{\Lambda}) \;=\;
   \int\int_0^{\infty} \, 
   \rg_{\,\beta}^{\,T_{\beta}} (r)\
   \mathrm{j}_{\Lambda} (k r)\
   \rg_{\,E\kappa}^{\,T \tilde{T}} \,(r,\, r')\
   \mathrm{j}_{\tilde{\Lambda}} (\tilde{k} r')\
   \rg_{\,\alpha}^{\,T_{\alpha}}(r')\:dr\,dr' \; , 
\end{equation}
where $\mathrm{j}_{\Lambda} (k r)$ denotes a spherical Bessel function,
$\,\rg_{E\kappa}^{\,T \tilde{T}} \,(r,\, r')\,$ one of the components of the 
radial 
Green's function in (\ref{rgf-first-equation}), and where $\alpha$ and $\beta$ 
refer to the radial wave functions of some bound or free--electron state, 
respectively. Matrix elements of this type play a key role, for instance, for
studying two--photon ionization (bound--free) or decay (bound--bound) 
processes. Since the (four) components of the radial Green's functions depend
on two radial coordinates, $r$ and $r'$, a rather large amount of data
need usually to be calculated by means of \textsc{Xgreens}. These data are 
finally stored in an external (\texttt{.rgf}) radial Green's function file 
as will be discussed in the next subsection.

\medskip

To generate and utilize the Green's functions, two derived data types
\texttt{TGreens\_single\_rgf} and \texttt{TGreens\_rgf} have been introduced 
as shown in Figure~\ref{f:data-structures}. Using these data types, a single
Green's function is kept internally in terms of its expansion coefficients
from Eqs.\ (\ref{piecewise-M}) and (\ref{piecewise-W}) for the
\textit{large--large} and a similar set of coefficients for the
\textit{small--small} component on the given grid. Apart from the particular
representation of the grid, these data structures contain the energy 
and symmetry of the radial Green's function, the nuclear charge function(s)
as well as information about the mode of interpolation of the radial 
components between the grid points and the maximum tabulation 
point \texttt{mtp}. In \textsc{Xgreens}, a variable of
\textsf{type(TGreens\_rgf)} is used in order to store the information about
all the requested functions, and a few additional procedures are provided 
to calculate from these structures the values of the radial components 
for any set of arguments.

\medskip

The basic steps in the execution of the program are very similar as for  
other components of the \textsc{Ratip} package. At the beginning, 
all the necessary input data are read in by an (interactive) dialog. 
The requested Green's functions are then calculated in turn and tested for 
their accuracy. The (main) output of the program is a formatted \textsc{Ascii} 
file which provides an interface for further applications, either within 
the \textsc{Ratip} environment or as worked out by the user.

\begin{figure}
{\renewcommand\baselinestretch{0.81}
\begin{footnotesize}
\begin{verbatim}   
   type ::  TGreens_single_rgf
   !-----------------------------------------------------------------
   !  Data structure type to keep a single Green's function
   !-----------------------------------------------------------------      
      real(dp)                              :: energy
      integer                               :: kappa
      ! Coefficients for the regular and irregular solutions
      real(dp), dimension(:,:), allocatable :: fLL1, fLL2, fSS1, fSS2
      ! Values of the regular and irregular solutions and their derivatives
      real(dp), dimension(:), allocatable   :: mL,  wL,  mS,  wS, &
                                               mLp, wLp, mSp, wSp
   end type TGreens_single_rgf
   !
   !
   type ::  TGreens_rgf
   !-----------------------------------------------------------------
   !  Data structure type to save the Green's functions
   !-----------------------------------------------------------------   
      type(TGreens_single_rgf), dimension(:), allocatable :: gf
      integer                             :: interpolation_mode, mtp
      real(dp), dimension(:), allocatable :: yp, z0, z1
   end type TGreens_rgf
\end{verbatim}
\end{footnotesize}}
\vspace*{-0.3cm}
\caption{\label{f:data-structures}\small The derived data structures 
\texttt{TGreens\_single\_rgf} and \texttt{TGreens\_rgf} to generate, store, 
and interpolate a central--field Green's function.}
\end{figure}

\subsection{Interactive control and output of the program}

Like the other components of the \textsc{Ratip} package, \textsc{Xgreens} is
controlled by a dialog at the beginning of the execution. In this dialog, 
all information need to be specified for obtaining the central--field 
potential as well as about the number and symmetry of the radial Green's 
functions to be generated. Owing to various choices for defining the spherical
potential, however, slightly different dialogs may occur during the execution. 
An example is shown in Figure~\ref{figure:2}, where first a 
Hartree--plus--statistical-exchange (HX) potential due to Cowan 
\cite{Cowan, Cowan:1967} is generated from the ground--state wave functions 
of atomic gold, and then two radial Green's functions are calculated within 
this potential.

\medskip

Apart from the nuclear charge, given in terms of a \textsc{Grasp92}
\texttt{.iso} isotope data file, the dialog first prompts for specifying 
(or confirming) some basic parameters such as the speed of light, the 
grid parameters as well as for the energy units, in which the further input 
and output is done by the program. This is followed by a prompt for
determining the central--field potential for the radial Green's functions.
At present, \textsc{Xgreens} supports four models including a pure Coulomb
potential (for the calculation of Coulomb Green's functions) as well
two potentials (\texttt{HFS} and \texttt{HX}) to incorporate also the 
\textit{exchange interaction} in some approximate form. In the \texttt{Hartree}
model, in contrast, only the \textit{direct} potential of some atomic level
is applied in the later computation of the Green's functions. In all these 
cases, however, the wave functions of the selected state have to be specified 
in terms of the corresponding \textsc{Grasp92} files for the configuration 
state functions (CSF), mixing coefficients as well as the radial orbitals. 
If computed internally, the potential can be written also to disc and used 
in further applications. Beside of the internal set--up of the potential, 
in addition, a user--defined potential can be given also explicitly and 
\textit{read in} by the program. 

\medskip

Although the nuclear charge is specified by means of a \texttt{.iso} 
file (in line with all other components of \textsc{Ratip}), the Green's 
functions are always generated for a \textit{point--like} nucleus in order
to 'utilize' their regular behaviour of the radial components at the origin
[cf.\ section \ref{ss:generation-of-rgf}]. For each radial Green's function,
then the energy and the one--particle symmetry $\kappa$ need to be specified.
The functions are finally written to the \texttt{.rgf} radial Green's function
file whose name has also to be given at the end of the dialog. In this file, 
the radial Green's functions are given at the grid (in $r$ and $r'$) as 
specified originally. Moreover, to test the accuracy of the Green's functions,
a number of overlap and normalization integrals are calculated on request 
in order to 'compare' a few low--lying orbitals, as obtained from the Green's 
functions by an integration over $\,r'$, with independently generated functions 
from Salvat's program \cite{Salvatetal:1995}. In average, the computation 
of a single Green's function requires about 2 min on a 450 MHz Pentium PC 
and approximately the same time for the 'test' integrals. In the latter case, 
the rather large demand on \textsc{CPU} time arises mainly from the
$2\,i_{\rm\, max}$ radial integrals in $r'-$space, which need first to be
performed in order to obtain the radial orbital functions from the given 
Green's function components.

\subsection{Distribution and installation}

A program of \textsc{Ratip}'s size cannot be maintained over a longer period
without that certain changes and the adaptation of the code to recent
developments (in either the hardware or the operating systems) become 
necessary from time to time. For the distribution of the code, therefore, 
we follow our previous style in that the \textsc{Ratip} package is provided 
as whole. The main emphasis with the present extension of the program, however,
is placed on the design and implementation of the two (new) modules 
\texttt{rabs\_greens.f90} and \texttt{rabs\_utilities\_2.f90} which contain 
the code for the generation and the test of the radial Green's functions.
In addition, the four modules (\texttt{rabs\_special\_functions.f90}, 
\texttt{rabs\_error\_control.f90}, \newline
\texttt{rabs\_greens\_external.f90}, and \texttt{rabs\_cpc\_salvat\_1995.f}) 
became necessary and had to be appended to the code 
in order to support an accurate computation of the hypergeometric and 
a few related functions from mathematical physics as well as for providing 
Salvat's solver \cite{Salvatetal:1995} for the generation of the radial
wave functions $(\rP(r),\,\rQ(r))$ within a given potential.

\medskip

In the present version, the \textsc{Ratip} package now contains 
the source code for the 8 components \textsc{Anco}, \textsc{Cesd}, 
\textsc{Lsj}, \textsc{Greens}, \textsc{Rcfp}, \textsc{Relci}, \textsc{Reos}
as well as the \textsc{Utilities} for performing a number of small but
frequently occurring tasks. Together with the six additional modules from above,
the overall program therefore comprises about 40.000 lines of code, separated
into 22 modules (apart from the main program components and three 
libraries). All these source files are provided in the
\texttt{Ratip} root directory.  As before, there is one \textit{makefile}
for each individual component from which the corresponding executable can be
obtained simply by typing the command \texttt{make -f make-component}, that is
\texttt{make -f make-greens} in the present case.
For most components of the \textsc{Ratip} package, in addition, we provide  
a test suite in a subdirectory \texttt{test-component} of the root where 
\texttt{component} refers to the names above. For example, the directory 
\texttt{test-greens} contains all the necessary files to run the sample
calculations which we will discuss below.

\medskip

Before, however, the makefiles can be 'utilized', a number of global variables
need to be specified for the compilation and linkage of the program.
In a \textsc{Linux} or \textsc{Unix} environment, this is achieved by 
modifying (and \textit{source}ing) the script file \texttt{make-environment} 
which saves the user from adopting each makefile independently. 
In fact, the script \texttt{make-environment} just contains a very few lines 
for specifying the local compiler, the options for the compiler, as well as 
the local paths for the libraries. Under \textsc{Windows}, in contrast,
not much help need to be given to the user since most compiler nowadays
provide the feature to define 'projects'. In this case, it is recommended to 
\textit{read off} the required modules and source files from the makefile 
of the program component and to 'declare' them directly to the project as
associated with the given component.

\medskip

In the past year, the \textsc{Xgreens} program has been applied under several
\textsc{Linux} systems using the \textsc{Lahey} Fortran 95 compiler. 
The \textsc{Ratip} program as a whole has been found also portable rather 
easily to other platforms
such as \textsc{IBM RS}/6000, \textsc{Sun OS}, or to the \textsc{PC} world.
The file \texttt{Read.me} file in the \texttt{ratip} root directory contains
further details for the installation. Overall, however, it is expected that it
will not be difficult to compile the program also under other operating systems.

\begin{figure}
\begin{scriptsize}
\vspace*{-1.5cm}

{\renewcommand\baselinestretch{0.86}
\begin{verbatim}
 XGREENS: Calculation of relativistic central-field Green's functions
  with energies E < 0   (Fortran 95 version);
  (C) Copyright by P Koval and S Fritzsche, Kassel (2004).

 Enter a file name for the greens.sum file:
z79-au-pot.sum
 Enter the name of the isotope data file:
z79.iso
 loading isotope data file ...
  ... load complete;
 The physical speed of light in atomic units is 137.0359895000000 ;
  revise this value ?
n
 The parameters of the radial grid 1 (first grid) are
  rnt =  2.177968408335618E-04 ;
  h   =  6.250000000000000E-02 ;
  hp  =  0.000000000000000E+00 ;
  n   =  390 ;
  revise these values ?
n
  Which units are to be used to enter and to print the energies:
    A       : Angstrom;
    eV      : electron volts;
    Hartree : Hartree atomic units;
    Hz      : Hertz;
    Kayser  : [cm**(-1)];
eV
  Determine type of central-field potential to generate:
    Load    : Load a potential from a .pot file
    Coulomb : (pure) Coulomb potential
    Hartree : Direct potential for a given ASF
    HFS     : Hartree-Fock-Slater method
    HX      : Hartree-plus-statistical-eXchange (Cowan 1967)
HX
 Enter the name of the .csl (configuration states list) file:
au.csl
 Loading configuration symmetry list file ...
  There are  22  relativistic subshells;
  there are  3  relativistic CSFs;
  ... load complete.
 Enter the name of the GRASP92 mixing coefficient file:
z79-au.mix
 Loading mixing coefficients file ...
  ... load complete;
 Enter the name of the rwf data file:
z79-au.rwf
  ... load complete;
  Summary of all ASF to facilitate the selection of  proper bound-state levels

     i   Level    J^P   Energy (a.u.)     Energy (eV)
  ----------------------------------------------------
     1    1    1/2 +   -1.90397E+04      -5.18096E+05

  Enter ASF level number for which the potential should be calculated:
1
  Enter the subshell (e.g. 1s, 2p-, ...) which is specific to Cowan's HX method:
1s
  Write the generated potential to disc ?
y
  Enter the name of .pot central-field potential file:
z79-au-hx.pot
 Radial Green's functions will be calculated within the range  0 <= r, r' <=    4.797    a.u.
  with the boundaries of the nuclear charge function Z(0) =    79.00     and Z(r_max) =   1.951
  In the given potential the one-electron energies (in eV) are:
   1s:  -8.131E+04
   2s:  -1.453E+04   2p-: -1.402E+04   2p:  -1.214E+04
   3s:  -3.479E+03   3p-: -3.232E+03   3p:  -2.814E+03   3d-: -2.382E+03   3d:  -2.289E+03

  Enter in turn the energies E (in eV) and symmetries of the radial Green's functions to be generated;
  Enter another energy and symmetry (<cr> if done):
-10000 s
  Enter another energy and symmetry (<cr> if done):
-15000 d-
  Enter another energy and symmetry (<cr> if done):

  Check the accuracy of the Green's functions after generation ?
y
  Enter the name of .rgf file to save the Green's function:
z79-au-pot.rgf
.
.
\end{verbatim} }
\end{scriptsize}
\vspace*{-0.4cm}
\caption{\label{figure:2}\small Interactive dialog for calculating the 
effective potential and for generating the radial Green's functions.}
\end{figure}

\begin{figure}
\begin{scriptsize}
{\renewcommand\baselinestretch{0.86}
\begin{verbatim}
 .
 . 
 The parameters of the radial grid 1 (first grid) are
  rnt =  2.177968408335618E-04 ;
  h   =  6.250000000000000E-02 ;
  hp  =  0.000000000000000E+00 ;
  n   =  390 ;
  revise these values ?
y
 Enter rnt:
2.177968408335618E-04
 Enter h:
3.125E-2
 enter hp:
0
 enter n:
780
  Which units are to be used to enter and to print the energies:
    A       : Angstrom;
    eV      : electron volts;
    Hartree : Hartree atomic units;
    Hz      : Hertz;
    Kayser  : [cm**(-1)];
eV
  Determine type of central-field potential to generate:
    Load    : Load a potential from a .pot file
    Coulomb : (pure) Coulomb potential
    Hartree : Direct potential for a given ASF
    HFS     : Hartree-Fock-Slater method
    HX      : Hartree-plus-statistical-eXchange (Cowan 1967)
Load
  Enter the name of a .pot file with a central-field potential
z79-au-hx.fld
 Radial Green's functions will be calculated within the range  0 <= r, r' <=    5.269    a.u.
  with the boundaries of the nuclear charge function Z(0) =    79.00     and Z(r_max) =   1.000
  In the given potential the one-electron energies (in eV) are:
    1s:  -8.131E+04
    2s:  -1.453E+04   2p-: -1.402E+04   2p:  -1.214E+04
    3s:  -3.479E+03   3p-: -3.232E+03   3p:  -2.814E+03   3d-: -2.382E+03   3d:  -2.290E+03

  Enter in turn the energies E (in eV) and symmetries of the radial Green's functions to be generated;
  Enter another energy and symmetry (<cr> if done):
-1000 s
  Enter another energy and symmetry (<cr> if done):
-10000 s
  Enter another energy and symmetry (<cr> if done):
-15000 p
  Enter another energy and symmetry (<cr> if done):
-10000 p-
  Enter another energy and symmetry (<cr> if done):
-15000 d-
  Enter another energy and symmetry (<cr> if done):
-10000 d
  Enter another energy and symmetry (<cr> if done):

  Check the accuracy of the Green's functions after generation ?
y
  Enter the name of .rgf file to save the Green's function:
z79-au-rgf.rgf
 .
 .  
\end{verbatim}}
\end{scriptsize}
\vspace*{-0.4cm}
\caption{\label{figure:3}\small Dialog for calculating the 
radial Green's functions with an increased number of radial points.}
\end{figure}

\section{\label{s:test-calculations}Test calculations}

To illustrate the use of the \textsc{Xgreens} program, we briefly discuss and
display two examples. They refer to the generation of the radial Green's
function for a few selected (one--electron) symmetries in atomic gold 
($Z \,=\, 79$) and demonstrate how the accuracy of these functions can be 
increased by a proper choice of the radial grid. However, before the radial 
components can be generated, we need first to specify the potential as shown 
in Figure~\ref{figure:2}. For the present examples, we have chosen a
\texttt{HX} (Hartree--plus--statistical--exchange) due to the work of Cowan
\cite{Cowan, Cowan:1967}, starting from the ground--state wave functions of
neutral gold. As usual for \textsc{Grasp92} and \textsc{Ratip}, these wave
functions have to be provided in terms of the \texttt{.csl} configuration state
list and the \texttt{.mix} configuration mixing files as well as the 
\texttt{.out} radial orbital file. All of these input files are provided also
in the \texttt{test-greens/} subdirectory of the \texttt{Ratip} root 
directory and, thus, can be used for the present tests. In Figure~\ref{figure:2}
moreover, the potential is saved to the central--field potential file 
\texttt{z79-au-hx.pot} and later re--utilized in Figure~\ref{figure:3}.
The output of the two examples are shown in the \textsc{Test Run Output} below.

\medskip

Like for the wave functions, the 'quality' of the generated Green's functions
becomes only fully apparent, if they are used for calculating 
\textit{observables} which can be compared to experiment. Although calculation 
of observables is not the aim of the present work, it is still possible to
test the accuracy of the generated Green's functions by re--calculating the
radial orbitals (as discussed in section \ref{ss:accuracy-test})
and by comparing these orbitals with those as obtained from the integration
of the Dirac equation for the given central field. This 'comparison' is done
in \textsc{Xgreens} on request by evaluating for the two lowest principal
quantum numbers $n$ the (radial) overlap integrals of the re--calculated
functions with the solutions by Salvat's program. In addition, the
\textit{normalization} integrals are also
displayed as seen from the \textsc{Test Run Output}. Except of a few cases,
the deviation of these integrals from $1.0$ are typically well below 0.1~\%{}
on the standard grid with about 300 mesh points in $r$ and $r'$. For these
deviations, there are two source of numerical 'inaccuracy' which arise from
the test procedure and are not related to the generation of the radial Green's
functions. They are caused by the fact that by using Salvat's solver 
 \cite{Salvatetal:1995}, a third--order spline is used to represent the
potential at intermediate points in $r$ in contrast to the linear
representation (\ref{piecewise-linear-approximation}) for the generation of 
the Green's functions. Moreover, the integration of the radial integrals 
uses the standard \textsc{Grasp92} procedure and, hence, is not adapted 
as well to Salvat's solutions.

\medskip 

The accuracy of the Green's functions (and the test integrals) can be 
improved however if the number of grid is enlarged, i.~e.\ the spacing between
the mesh points reduced. In our second example, cf.\ Figure~\ref{figure:3}, we
therefore adopt a grid with approximately twice the number of mesh points.
Note that this requires four times as much storage as the radial components
are functions in $r$ and $r'$. In this test case, moreover, the radial Green's
functions are generated for four different symmetries 
$s,\, p_{1/2},\,  p_{3/2},\,  d_{3/2}$, and  $d_{5/2}$. When the 
\textsc{Test Run Output} from this example is compared with those from
example 1, we see that the accuracy in the overlap integrals in increased by
about a factor of 5. For this example, the generation and test took about 
30 minutes in total on a 450 MHz Pentium III.

\medskip

After the termination of \textsc{Xgreens}, the radial Green's functions file
(e.~g.\ \texttt{z79-au-rgf.rgf} from example 2) contains all the Green's
function components which were generated during the execution. Since this is a 
formatted file, it can easily be read and manipulated by any text editor.
This (radial Green's function) file stores the individual components
$\rg_{E\kappa}^{\,TT'}\,(r,\,r')$ of the Green's functions as 2--dimensional
arrays within a simple file structure. An (internal) file signature 
\texttt{\# DCFGF} in the first line is followed by the mode of interpolation
and the number of Green's functions which are provided by the file.
Then, each Green's function is specified (in turn), starting with the
energy and symmetry of the function and followed by a table of six rows where
the first two rows refer to the coordinates $r$ and $r'$ (in atomic units) and
and the other rows to the four component functions 
$\rg_{E\kappa}^{\,LL}(r,r')$,    $\rg_{E\kappa}^{\,LS}(r,r')$,
$\rg_{E\kappa}^{\,SL}(r,r')$ and $\rg_{E\kappa}^{\,SS}(r,r')$ in this 
particular order. Although this format is convenient for the later use of
these functions, it may lead to rather large data files, especially if
some larger number of Greens' functions need to be generated. In practise, 
however, not much problems are expected with this file structure because disc
storage became cheap recently and because, for most applications, we expect the
Green's functions to be generated \textit{on demand} by making \texttt{use} 
of the corresponding modules instead of keeping them in external files.
Typically, the radial wave and Green's functions occur as part of matrix
elements and, thus, first require an additional integration (over $r$ 
and/or $r'$) before any \textit{observable} quantity is obtained. 

\medskip

\begin{figure}
\vspace*{-2.5cm}
\epsfig{file=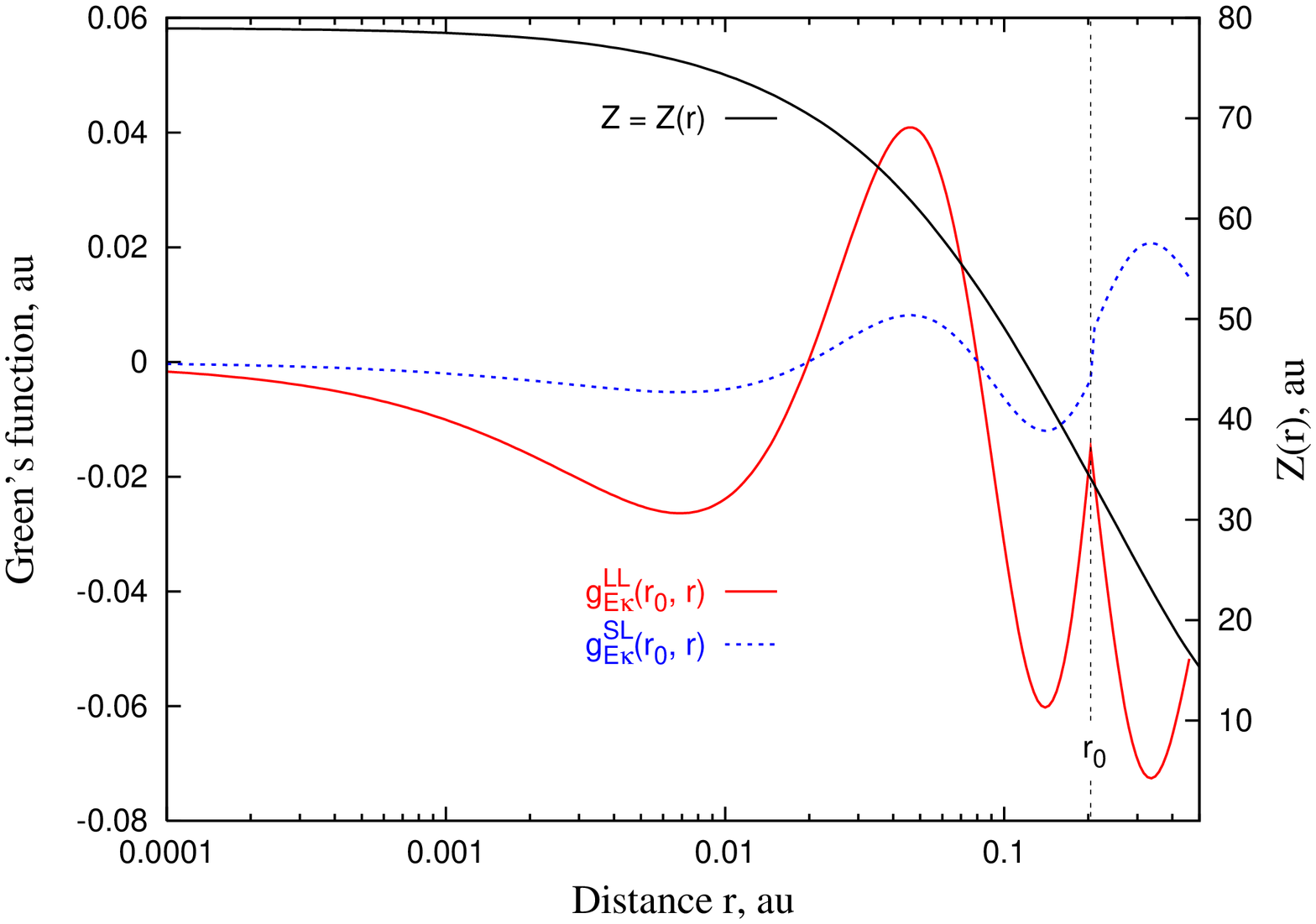, width=11.0cm, angle=270}
\caption{\small Nuclear charge $\rZ\,=\,\rZ(r)$ and the radial Green's 
function 
components $\rg^{\,LL}_{E\,s}\,(r_0, r)$ and $\rg^{\,SL}_{E\,s}(r_0, r)$
as function of $r$ and for $r_0 \,=\, 0.204076$ a.u. The radial 
Green's functions with $s-$symmetry ($\kappa \,=\, -1$) are calculated 
in the ground--state potential of atomic gold for the energy 
$E\,= \, 1000$ eV.} 
\label{f:inout}
\end{figure}

The format of the radial Green's function (\texttt{.rgf}) file can be used
easily in order to plot the various radial components. In
Figure~\ref{f:inout}, therefore, we display the nuclear charge function
$\rZ(r)$ and the two radial Green's function components 
$\rg^{\,LL}_{E\,s}\,(r_0, r)$ and $\rg^{\,SL}_{E\,s}(r_0, r)$ as function of 
(second) argument $r$ and taken for $r_0 \,=\, 0.204076$ a.~u.
The radial Green's functions are calculated in the ground--state potential of
atomic gold, as applied in example 2, for the energy $E\,= \, 1000$ eV.
As seen from Figure~\ref{f:inout}, the \textit{large--large} component
$\rg^{\,LL}_{E\,s}$ is continuous at $r\,=\,r'$ while the \textit{small--large}
component $\rg^{\,SL}_{E\,s}$ jumps owing to Eqs.\  (\ref{jump}) and 
(\ref{jump-SL}) at this position in space. Both components, moreover, behave
'regularly' at the origin as was constructed explicity by Eq.\
(\ref{product-M-W}).

\section{\label{s:summary-and-outlook}Outlook. Applications of central--field 
         Green's functions}

Apart from fundamental interest in having the central--field Green's functions
available for 'relativistic' electrons, these functions are also useful for 
the perturbative treatment of atoms and ions. As discussed in subsection
\ref{ss:accuracy-test},
namely, the Green's function provide a simple access to the complete spectrum
of a quantum system and, hence, can be utilized for carrying out the summation 
over all the (one--particle) states of the spectrum as required in 
second-- and higher--order perturbation theory. However, since we only provide
the \textit{one--electron} Green's functions, they can be applied for just 
those processes where the 'perturbations' are described in terms of
one--particle operators. Perhaps, the most--studied perturbation of this 
type is the interaction of atoms and ions with the radiation field,
which --- within a sufficiently strong field ---  may lead not only to atomic
photoionization but (in second order) also to the two--photon ionization and 
decay of atoms and ions and to several other processes. During the last two
years, we utilized the central--field Green's functions mainly for exploring
the two--photon ionization for the helium--like ions and for the
inner--shell electrons from atomic neon and argon.

\medskip

Apart from the one--particle Green's functions (as discussed in this work), 
there are a number of further processes such as the double Auger decay, 
for which the \textit{two--particle} Green's functions 
$\:\rG_E(\br_1,\br_2;\,\br_1',\br_2')\:$
are required in order to calculate the cross sections, decay rates 
and/or angular distributions. These two--particle functions are solutions 
to a \textit{defining equation} similar to Eq.\  
(\ref{CoulombDiracGreensEquation}) but where the (one--particle) Dirac 
Hamiltonian $\hat{H}_{\rm\, D}$ is replaced by the two--particle operator 
$\hat{H}_{\rm\, D,1} \,+\, \hat{H}_{\rm\, D,2} \,+\, \hat{V}_{12}$,
including the electron--electron interaction $\hat{V}_{12}$ explicitly. Even by 
making use of a proper \textit{radial--angular} decomposition of such Green's 
functions,
their radial part  $\rg_{\,\kappa_1,\kappa_2}^{\,k} (r_1,r_2; r_1',r_2')$ 
would depend then on four radial variables and an overall rank $k$,
similar as known from the tensorial decomposition of the electron--electron 
interaction \cite{Varshalovich}. Therefore, an internal representation and 
generation of these (radial) functions appear rather infeasible in practise. 
An alternative to these two--particle Green's functions is given by the 
(so--called) \textit{modified pair functions} $\Phi_{E,\alpha} (\br_1,\br_2)$
which satisfy the equation
$$
   \left\{ \hat{H}_{\rm\, D,1} \,+ \,\hat{H}_{\rm\, D,2} \,+ \, 
      \hat{V}_{12}\,-\,E \: \right\}\, \Phi_{E,\alpha} (\br_1,\br_2) \;=\; 
      \hat{W} \, \left| \Psi_\alpha (PJM) \right\rangle \, ,
$$
where $\hat{W}$ is a one-- or two--particle operator which depends on the
physical task to be solved and $\left| \Psi_\alpha (PJM) \right\rangle$
is one of the (two--electron) solutions of the corresponding homogeneous
equation. These modified pair functions have the advantage that they only
depend on two radial variables similar as the one--particle Green's functions.
On the other hand, of course, these functions now depend on some (initial)
two--particle state $\left| \Psi_\alpha (PJM) \right\rangle$ as well as on 
the particular choice of the interaction operator $\hat{W}$ and, hence, 
are less general when compared with the Green's function from above. 
We currently investigate the possibilities for generating also such 
modified pair functions within the framework of \textsc{Ratip} and for 
their (later) use in the description of the double Auger decay. 
But already the one--particle Green's functions from this work enables 
one to explore a number of atomic properties which have not been studied 
before for complex atoms or, at least, not within a relativistic framework.

\bigskip

\subsubsection*{Acknowledgement:}

This work has been supported by the Deutsche Forschungsgemeinschaft (DFG) 
under the contract Fr 1251/8--2 and in the framework of the 
\textit{Schwerpunkt} SPP 1145.

\bigskip

{\renewcommand\baselinestretch{0.95}
\begin{small}

\end{small} }

\newpage        
\section*{TEST RUN OUTPUT} 

\begin{scriptsize}
{\renewcommand\baselinestretch{0.86}
\begin{verbatim}
***************
** Example 1 **
***************

  Generation of radial Green's functions

     i     E (       eV)    j     overall progress
  ---------------------------------------------------
     1    -1.0000000E+04    s         50%
     2    -1.5000000E+04    d-       100%

  
  Tests on the accuracy of the Green's functions by means of overlap and normalization integrals:

                                      Overlap integrals          Normalization
     i     E (       eV)   nj  <nj (Greens) | nj (Salvat 95)>   ||nj (Greens)||
  -------------------------------------------------------------------------------
     1    -1.0000000E+04   1s          1.000020E+00              1.000057E+00
     1    -1.0000000E+04   2s          1.001925E+00              1.003888E+00
     2    -1.5000000E+04   3d-         9.997536E-01              9.995404E-01
     2    -1.5000000E+04   4d-         1.005937E+00              1.012284E+00
  -------------------------------------------------------------------------------


 Write the Green's functions to the .rgf file;
  2 radial Green's functions with 5065446 bytes ...

 XGREENS complete ...
\end{verbatim}}
\end{scriptsize}

\vspace*{1.0cm}

\begin{scriptsize}
{\renewcommand\baselinestretch{0.86}
\begin{verbatim}
***************
** Example 2 **
***************

  Generation of radial Green's functions

     i     E (       eV)    j     overall progress
  ---------------------------------------------------
     1    -1.0000000E+03    s         16%
     2    -1.0000000E+04    s         33%
     3    -1.5000000E+04    p         50%
     4    -1.0000000E+04    p-        66%
     5    -1.5000000E+04    d-        83%
     6    -1.0000000E+04    d        100%


  Tests on the accuracy of the Green's functions by means of overlap and normalization integrals:

                                      Overlap integrals          Normalization
     i     E (       eV)   nj  <nj (Greens) | nj (Salvat 95)>   ||nj (Greens)||
  -------------------------------------------------------------------------------
     1    -1.0000000E+03   1s          1.000267E+00              1.000540E+00
     1    -1.0000000E+03   2s          1.000318E+00              1.000641E+00
     2    -1.0000000E+04   1s          1.000003E+00              1.000010E+00
     2    -1.0000000E+04   2s          1.000138E+00              1.000278E+00
     3    -1.5000000E+04   2p          9.993563E-01              9.987343E-01
     3    -1.5000000E+04   3p          9.999064E-01              9.998404E-01
     4    -1.0000000E+04   2p-         1.000384E+00              1.000770E+00
     4    -1.0000000E+04   3p-         1.000216E+00              1.000435E+00
     5    -1.5000000E+04   3d-         1.000046E+00              1.000107E+00
     5    -1.5000000E+04   4d-         1.001539E+00              1.003181E+00
     6    -1.0000000E+04   3d          1.000084E+00              1.000173E+00
     6    -1.0000000E+04   4d          1.001030E+00              1.002106E+00
  -------------------------------------------------------------------------------


 Write the Green's functions to the .rgf file;
  6 radial Green's functions with 59754187 bytes ...

 XGREENS complete ...
\end{verbatim}}
\end{scriptsize}

\end{document}